# Towards a Group Theoretic Quantum Encryption Scheme Based on Generalized Hidden Subgroup Problem.

Srinivasan N, Sanjeevakumar C, Sudarsan L, Kasi Rajan M, Venkatesh R
Department of Information Technology,
Madras Institute of Technology,
Chennai 600044
Email: ns@annauniv.edu

**Abstract:**

*This paper introduces a completely new approach to encryption based on group theoretic quantum framework. Quantum cryptography has essentially focused only on key distribution and proceeded with classical encryption algorithm with the generated key. Here, we present a first step towards a quantum encryption scheme based on the solution for the hidden subgroup problem. The shared secret key K from QKD evolves as a generator for a subgroup H of a group G, in combination of the plain text data modeled as group elements. The key K helps in regeneration of the plain data on the receiver's side based on subgroup reconstruction. This paper models all quantum computations using group representations. A non-constructive proof is attempted towards the security of the encryption scheme. We also address the issues involved in a such a venture into the realms of Quantum data encryption.*

# 1. Introduction

Until now, quantum cryptography has essentially concentrated on Quantum key Distribution. Many key distribution schemes have been proposed based on the BB84 protocol. This paper is an extension of our work in QKD, exploiting the randomness of measurement bases and key values to reduce unauthorized information gain by an eavesdropper [5] during key distribution. Unconditional security has been achieved based on the principles of Quantum Theory. The encryption algorithm used is however classical. The necessity for a quantum encryption algorithm would arise in the face of availability of immense computing power in the system which includes communicating parties and eavesdroppers.

This paper is probably the first attempt towards an encryption scheme exploiting quantum principles. The approach is based on the Hidden Subgroup problem. The importance of the Hidden Subgroup Problem [1, 2] is that it encompasses most of the quantum algorithms found so far that are exponentially faster than their classical counterparts. Research in this area is centered on extending the families of groups for which the HSP can be efficiently solved, which may improve other classically inefficient algorithms, such as determining graph isomorphism or finding the shortest vector in a lattice. Finally, there are many group theoretic algorithms that are more efficient on a quantum computer, such as finding the order of a finite group given a set of generators.

There are lots of mathematical issues that need to be addressed with such an approach. The first consideration would be the dynamic construction of groups of abelian, non-abelian or normal nature. The inherent difficulty is modeling the key K as related to the generator of a subgroup. These issues require group theoretic explorations, which we have tried to present in a formal way. Section 2 of the paper gives a formal definition of the Hidden Subgroup Problem. The solution to the problem with respect to different groups is explored. A basic theoretical model of our proposal is detailed in Section 3 of the paper. Section 4 presents a non-constructive proof of security. The subsequent sections give the mathematical requirements for our proposal.

## 2. The Hidden Subgroup Problem

This section explores the literature for the Hidden Subgroup problem, in view of developing an equivalent quantum algorithm. We make a general definition of the Hidden Subgroup Problem, which we will abbreviate HSP for the rest of this paper, and then summarize for which groups G and subgroups H the HSP can be solved efficiently. We will also discuss partial results which make a general definition of the Hidden Subgroup Problem, which we will abbreviate HSP for the rest of this paper, and then attempt to determine for which groups G and subgroups H we can solve the HSP efficiently. We will also discuss partial results on groups for which efficient HSP algorithms are not known.

*Definition* (Separates cosets). Given a group G, a subgroup $H \leq G$ and a set X, we say a function f: G ->X separates cosets of H if for all $g_1, g_2 \in G$  f ($g_1$) = f ($g_2$) if and only if $g_1H = g_2H$.

*Definition* (The Hidden Subgroup Problem). Let G be a group, X a finite set, and f : G -> X a function such that there exists a subgroup H < G for which f separates cosets of H. Using information gained from evaluations of f, determine a generating set for H. For any finite group G, a classical algorithm can call a routine evaluating f(g) once for each $g \in G$ and thus determine H with O(|G|) function calls. A central challenge of quantum computing is to reduce this naive O(|G|) time algorithm to O(poly(log |G|)) time (including oracle calls and any needed classical post processing time). This can be done for many groups, which gives the exponential speedup found in most quantum algorithms.

### 2.1. The Solution for the Abelian Case.

The Solution [1, 2, 6] for the Abelian case has an equivalent experimental setup.

*a. Prepare the state*

$$\frac{1}{\sqrt{|G|}} \sum_{g \in G} |g, f(g)\rangle$$

and measure the second register f (g). As f takes distinct values on the left cosets of H, the resulting state is

$$\frac{1}{\sqrt{|H|}} \sum_{h \in H} |ch, f(ch)\rangle$$

where c is an element of G selected uniformly at random.

*b. Compute the Fourier transform of the coset state (1), resulting in*

$$\sum_{\rho \in G} \sqrt{\frac{1}{|G|}} \sqrt{\frac{1}{|H|}} \sum_{h \in H} \rho(ch) |\rho\rangle$$

where $\hat{G}$ denotes the set of homomorphisms $\{\rho: G \to C\}$

*c. Measure the first register and observe a homomorphism $\rho$*

A key fact about this procedure is that the resulting distribution over $\hat{G}$ is independent of the coset cH arising after the first stage (as the support of the first register in (1)). Thus, repetitions of this experiment result in the same distribution over $\hat{G}$

## 2.2 The Solution for the Non-Abelian Case.

A quantum algorithm which solves the hidden subgroup problem for the non-abelian groups [3] $W_n$ uses O(n) evaluations of the black box quantum circuit f and the classical post-computation, which is essentially linear algebra over $F_2$, also takes a number of operations which is polynomial in n.

*a. Prepare the ground state*

$$|\psi_1\rangle = |0...0\rangle \otimes |0...0\rangle$$ in both registers.

*b. Achieve equal amplitude distribution in the first register, for instance by an application of a Hadamard transform to each qubit:*

$$|\psi_2\rangle = \sum_{x \in W_n} |x\rangle \otimes |0...0\rangle$$

(Normalization factors omitted.)

*c. Calculate f in superposition and obtain*

$$|\psi_3\rangle = \sum_{x \in W_n} |x\rangle |f(x)\rangle$$

*d. Measure the second register and obtain a certain value z in the image of f. In the first register we have a whole coset $g_0U$ of the hidden subgroup U:*

$$|\psi_4\rangle = \sum_{f(x)=z} |x\rangle |z\rangle = \sum_{x \in g_0 U} |x\rangle |z\rangle$$

*e. Now solve the hidden subgroup problem for the normal subgroup N, which is the base group of $W_n$. This can be done by application of the standard algorithm for $Z_2^{2n}$ on the first 2n qubits.*

*f. Application of the Fourier transform on the first register using the circuit given in section 5 transforms the coset into a superposition of the form*

$$\sum_{x \in U^\perp} \psi_{g_0, y} |y\rangle$$

in case $g_0$ element of N (with certain phase factors $\psi_{g_0, y}$ which depend on $g_0$ and y and are from $\{\pm 1\}$). If $g_0$ element of $W_n \backslash N$ we get a superposition over the conjugated group

$$\sum_{x \in (U^t)^\perp} \psi_{g_0, y} |y\rangle$$

*g. Now measure the first register. With probability 1/2 we draw $g_0$ from N resp. $W_n \backslash N$, i.e., we get a superposition over $U^\perp$ resp. $(U^t)^\perp$ which leads (by performing measurements) to either equal distribution over $U^\perp$ or equal distribution over $(U^t)^\perp$*

*h. Iterating steps a-g* we generate with high probability the group $U^\perp \cap N$ and the group $(U^t)^\perp \cap N$.

What is missing are the sets $(U^\perp \cap N)t_0$ and $((U^t)^\perp \cap N)t_0'$ with certain elements $t_0$ and $t_0'$ not in N. It is clear that it is sufficient to find only one element in one of these two sets, since then the whole group $\langle U^\perp, (U^t)^\perp \rangle$ will be generated. But if any, there are many elements of this form in $U^\perp$ resp. $(U^t)^\perp$ since either there are none of them or exactly half of the elements of $U^\perp$ resp. $(U^t)^\perp$ is not in $U^\perp \cap N$ resp. $(U^t)^\perp \cap N$

Summarizing:

After performing this experiment an expected number of 4n times we generate with probability greater than 1-$2^n$ the group $\langle U^\perp, (U^t)^\perp \rangle$

*i. By solving linear equations* over F2 it is easy to find generators for

$$\left(\langle U^\perp, (U^t)^\perp \rangle\right)^\perp = U \cap U^t$$

After all we get generators for

$$U = (U \cap N)(U \cap U^t)$$

## 3. A Basic Introduction to Group Theoretic Quantum Encryption

This section describes the theoretic model of our approach towards an algorithm for encryption based on quantum principles. The most important axioms of quantum theory that makes this approach feasible are: Quantum Super-position and Measurement theory. All quantum algorithms have essentially been generalized to solve pattern identification, referred to as the solution for the Hidden Subgroup problem. With a formal definition of HSP already presented, we now present the details of our Quantum Encryption Protocol (QEP).

The first step requires establishing a relation between the key K generated from Quantum Key Distribution and the generator element 'g' of a subgroup of an arbitrary group to be constructed dynamically from plain text data (modeled as group elements).

An efficient relation would be is to consider K as the generator g of the subgroup. This results in a complexity in the form of reducing all plain text elements to the key K. In a real time setup of Quantum Key Distribution, K may not be related to the plain text data at all. However, in a classical key distribution scheme, K can be derived out of the plain text to be sent, and then K can be communicated to the other end of the channel. In our proposal we model every data as arbitrary elements that given a relation form a group dynamically. At the initiator's side (say Alice), the key K, in combination with plain text elements, is used to construct a subgroup H of a group G. The entire group is transmitted as the cipher text. At the receiver's side (say Bob), the subgroup of plain text elements is reconstructed using K.

## 4. Non-Constructive Proof of Security.

Now, an essential argument lies in the fact that an eavesdropping strategy by Eve can reconstruct the subgroup based on the solutions to HSP on different groups. This requires that we retort to a non-constructive proof of security of our proposal. Later, we present the mathematical details of the possibility of such a proposal. Given an arbitrary group, identifying the right subgroup of plain text elements is a probability. Eve is not aware of K or has very less information about it. This is one fact that ensures that Eve may not even recognize the right generator on solving HSP on the cipher data. One of the most important issues concerning the security of our proposal is the structure of the dynamically generated group G. For Eve to gain any relevant information on plain text data or K, she has to know the structure of G. We assume that G is abelian. There are two possible ways of computing the structure of the group. The first requires that the order of the abelian group is known, or that it can be computed. The second method does not require knowledge of the order of the group: the structure is computed from a set of generators. In this case, however, Eve happens to be ignorant of both details.

## 5. Quantum Encryption: Finite Group Construction

Our proposal requires that the order of the group G be dynamically fixed by Alice [4, 7, 8].
Let O(G) be n. Note that the value of n is immaterial to Bob as he reconstructs only the right subgroup containing the plain text data. We restrict ourselves to the construction of generic abelian group for the purpose. Note that the entire group would constitute the randomizer of the plain text data elements. We simply construct G and then construct compatible subgroup H containing plain text elements to be conveyed, with K as the generator element of H.

### 5.1. Construction of a Generic Abelian Group

A generic abelian group can be defined over any domain, universe, or aggregate. Alice may define the identity, the composition and the inverse functions. It is possible to determine if the elements of the group will be represented internally as a linear combination of the generators. To aid in the computation of the group, it is also possible to set the order and the generators of the group. Creating a generic abelian group does not imply that the group structure will be computed, unless this is explicitly required. However, structure computation will be automatically triggered when this is required by the nature of the access functions relative to the generic group created. Typically, one reason for not computing the structure of the group is that it may be expensive to do so and may actually not be required for further operations such as finding the order or the discrete logarithm of an element of the group.

We adopt an algorithmic approach to the construction of G. We introduce certain parameters for the purpose-IdIntrinsic, AddIntrinsic, InverseIntrinsic, UseRepresentation, Order, UserGenerators, ProperSubset, RandomIntrinsic, ComputeStructure.

Construct the generic abelian group G over the domain U. The domain U can be an aggregate (set, indexed set or sequence) of elements or it can be any structure, for example, an elliptic curve, a jacobian, a magma of quadratic forms. It can even be a set of unintelligible text elements that can distract the structure of the plain text elements.

If the parameters IdIntrinsic, AddIntrinsic and InverseIntrinsic are not set, then the identity, the composition and the inverse function of G are taken to be the identity, the composition and the inverse function of U or of Universe(U) if U is an aggregate. If the parameters IdIntrinsic, AddIntrinsic and

InverseIntrinsic are set, they define the identity, the composition and the inverse function of G. The parameter IdIntrinsic must be a function name whose sole argument is U or Universe(U) if U is an aggregate. AddIntrinsic must be a function name whose only two arguments are elements of U or of Universe (U) if U is an aggregate. InverseIntrinsic must be a function name whose only two arguments are elements of U or of Universe(U) if U is an aggregate. That is, it is required that InverseIntrinsic be a binary operation. Defining any of the three above parameters implies that the other two must be defined as well. Setting the parameter UseRepresentation to true implies that all elements of G will be internally represented as linear combinations of the generators of G obtained while computing the structure of G. This can be a costly procedure, since such a representation is akin to solving the discrete logarithm problem. The advantage of such a representation is that arithmetic for elements of G as well as computing the order of elements of G are then essentially trivial operations. The value of the parameter Order can be set to the order of the group, if it is known. This may be useful as it removes the need to compute the order of the group, should this be required by some group structure computation, or to solve the discrete logarithm problem. More importantly, if G is a proper subset of U, or of Universe(U) if U is an aggregate, then Order must be set, unless the parameter UserGenerators is set. One can set UserGenerators to some set of elements of U, or of Universe(U) if U is an aggregate, which generate the group G. This can be useful when G is a subset of U (Universe(U)), or more generally when the computation of the group structure of G is made from a set of generators. Finally, setting UserGenerators may be an alternative to setting RandomIntrinsic. The parameter ProperSubset must be set when G is a proper subset of U (Universe(U)). The parameter RandomIntrinsic indicates the random function to use. If it is not set, the random function (if it is required) is taken to be the random function applying to U (Universe(U)). The parameter RandomIntrinsic must be the name of a function taking as its sole argument U (Universe(U)) and returning an element of U (Universe(U)) which is also an element of the group G, which is important when G is a proper subset of U (Universe(U). Therefore, if G is a proper subset of U (Universe(U)), then RandomIntrinsic must be set, unless the parameter UserGenerators is set. The parameter ComputeStructure enables Alice to request that the computation of the group structure be performed at the time of creation.

### 5.2. Construction of the Subgroup

The same dichotomy between creation and group structure computation as the one discussed for the construction of G applies for subgroups of generic abelian groups. That is, if H is a subgroup of the generic abelian group G, then, as a general rule, the structure of H is not computed at creation time. There are two exceptions to this rule: when Alice specifically requests structure computation to be performed (via the ComputeStructure parameter), and when the structure of G is already known and H is given in terms of a set of generators (in our case, it is just K), or H is a p-Sylow subgroup of G. Construct the subgroup of G generated by the elements specified by the terms of the *generator list* L (in this case, it is K). A term L[i] of the generator list may consist of any of the following objects: an element liftable into G; a sequence of integers representing an element of G; a set or sequence whose elements may be of either of the above types. An element liftable into G may be an element of G itself, or it may be an element of U (Universe(U)), U being as usual the domain over which G is defined. The subgroup construction must be consistent with that of G.

In particular, it is possible to construct a subgroup by giving its order and a random function generating elements of the subgroup. Also, where the group structure of G is already known, or has been computed, and if the subgroup is defined in terms of K then the subgroup structure is computed at the time of creation.

# 6. Quantum Decryption of Authenticated Elements: Subgroup Reconstruction and Verification

This approach has its advantage in terms of incorporating implicit message authentication. The method of decryption is simply to reconstruct the subgroup with generator K. If G is soluble, then most of its subgroups are normal. There are efficient results for reconstruction of normal subgroups. Solving HSP on G can be used to cross check the authenticity of the plain text elements. A generic quantum algorithm for reconstruction [8] would be as follows

Compute

$$\sum_{g \in G} |g, f(g)\rangle$$

and measure the second register f(g).

The resulting super-position is

$$\sum_{h \in H} |ch\rangle \otimes |f(ch)\rangle$$

for some coset cH of H. Furthermore, c is uniformly distributed over G.

Compute the Fourier transform of the coset state, which is

$$\sum_{\rho \in \hat{G}} \sqrt{\frac{d_\rho}{|G|}} \sqrt{\frac{1}{|H|}} \left( \sum_{h \in H} \rho(ch) \right)_{i,j} |\rho, i, j\rangle$$

where $\hat{G}$ denotes the set of irreducible representations of G. Measure the first register and observe a representation on ρ. The probability of measuring the representation ρ is proportional to the dimension of ρ and number of times ρ appears in the induced representation $\text{Ind}^G_H 1_H$, where $1_H$ denotes the trivial representation on H.

# 7. Conclusion

Our original group theoretic approach opens up new dimensions to quantum cryptography, beyond key distribution. Most of the theoretic assumptions with respect to the relation between key K and plain text elements, properties of the generated groups and reconstruction can be relaxed with the advent of efficient solutions to HSP on arbitrary groups. We have modeled all data as group elements thereby elevating them to an abstract plane, beyond type and operation specifications. This approach will certainly trigger future research into quantum algorithms for encryption.